\documentclass[letterpaper,conference]{IEEEtran}

\usepackage{color}
\usepackage{caption}
\usepackage{subfig}
\usepackage{xfrac}
\usepackage{amsmath}
\usepackage{amsthm}
\usepackage{amssymb}
\usepackage{amsfonts}
\usepackage{relsize}
\usepackage{multirow}
\usepackage{threeparttable}

\usepackage{tikz}
\usetikzlibrary{shapes}
\usetikzlibrary{matrix}
\usetikzlibrary{arrows}
\usetikzlibrary{decorations.markings}
\usetikzlibrary{spy}
\usetikzlibrary{backgrounds}

\usepackage{pgfplots}
\usepgfplotslibrary{units}

\pgfplotsset{
    legend image with text/.style={
        legend image code/.code={%
            \node[anchor=center] at (0.3cm,0cm) {#1};
        }
    },
}

\tikzstyle{vecArrow} = [thick, decoration={markings,mark=at position
   1 with {\arrow[semithick]{open triangle 60}}},
   double distance=2pt, shorten >= 5.5pt,
   preaction = {decorate},
   postaction = {draw,line width=1.4pt, white,shorten >= 4.5pt}]

\definecolor{Set1-4-1}{RGB}{228,26,28}
\definecolor{Set1-4-2}{RGB}{55,126,184}
\definecolor{Set1-4-3}{RGB}{77,175,74}
\definecolor{Set1-4-4}{RGB}{152,78,163}
\definecolor{Set1-5-5}{RGB}{255,127,0}
\definecolor{Set1-7-1}{RGB}{228,26,28}
\definecolor{Set1-7-2}{RGB}{55,126,184}
\definecolor{Set1-7-3}{RGB}{77,175,74}
\definecolor{Set1-7-4}{RGB}{152,78,163}
\definecolor{Set1-7-5}{RGB}{255,127,0}
\definecolor{Set1-7-6}{RGB}{166,86,40}
\definecolor{Set1-7-7}{RGB}{0,0,0}

\newcommand{\figurewidth}{1}	
\newcommand{\doublefigurewidth}{0.55}	
\newcommand{\figureheight}{0.85}	
\newcommand{\ferfigureheight}{0.8}	

\tikzset{
leavenode/.style = {align=center, 
                    inner sep=2pt, 
                    text centered
                   },
imnode/.style = {align=center, 
                 inner sep=1pt, 
                 text centered, 
                },
fplus/.style={semithick,->,solid},
fminus/.style={semithick,->,dashed},
enc/.style={semithick,<-,solid},
factor/.style={semithick,<->,solid},
}

\title{Comparison of Polar Decoders with Existing Low-Density Parity-Check and Turbo Decoders}
\author{Alexios Balatsoukas-Stimming, Pascal Giard, and Andreas Burg\\Telecommunications Circuits Laboratory, EPFL, Switzerland\\%
Email: \{alexios.balatsoukas,pascal.giard,andreas.burg\}@epfl.ch}
\date{\today}

\usepackage{ifpdf}
\ifpdf
\usepackage[draft,pdfborder={0 0 0}]{hyperref}
\fi

\usepackage{balance}
\IEEEoverridecommandlockouts 

\begin{document}
\maketitle%
\begin{abstract}
Polar codes are a recently proposed family of provably capacity-achieving error-correction codes that received a lot of attention. While their theoretical properties render them interesting, their practicality compared to other types of codes has not been thoroughly studied. Towards this end, in this paper, we perform a comparison of polar decoders against LDPC and Turbo decoders that are used in existing communications standards. More specifically, we compare both the error-correction performance and the hardware efficiency of the corresponding hardware implementations. This comparison enables us to identify applications where polar codes are superior to existing error-correction coding solutions as well as to determine the most promising research direction in terms of the hardware implementation of polar decoders.
\end{abstract}

\section{Introduction}
Polar codes~\cite{Arikan2009} are a new class of provably capacity-achieving channel codes which have attracted significant attention due to their interesting theoretical properties and their low-complexity encoding and decoding algorithms. The most popular decoding algorithms for polar codes are simple successive-cancellation (SC) decoding~\cite{Arikan2009}, successive-cancellation list (SCL) decoding~\cite{Tal2011}, and belief-propagation (BP) decoding~\cite{Pamuk2011}. 

Polar codes are currently under consideration for potential adoption in future 5G standards. A crucial factor in this discussion is the demonstration of polar decoder ASIC implementations that can outperform existing decoders for error-correction codes. A comparison against low-density parity-check (LDPC) decoders and Turbo decoders, both in terms of the error-correction performance and in terms of the hardware efficiency, is of particular interest. We note that other decoder properties, such as the flexibility, the decoding latency, and the energy efficiency, are also of great importance~\cite{Kienle2011}, but they are beyond the scope of this paper.

Over the past few years, significant advances have been achieved in the hardware implementation of decoders for polar codes. A brief overview of the most important ones can be found in~\cite{Giard2016c}. Error-correction performance comparisons of polar decoders against that of decoders for other error-correction codes can sporadically already be found in the literature. For example,~\cite{Pamuk2011} compared an FPGA-based BP polar decoder with an FPGA-based decoder for the Turbo code of the IEEE~802.16e standard. Moreover,~\cite{Sarkis2014} compared an FPGA-based SC decoder with an FPGA-based decoder for the LDPC code of the IEEE 802.3an standard. The authors of~\cite{Tal2015} compared the error-correction performance of SCL decoding with the error-correction performance of the LDPC code used in the IEEE 802.16e standard. Finally,~\cite{Sarkis2016} compared SCL decoding with the LDPC codes used in the IEEE~802.11n and IEEE~802.3an standards. However, these comparisons were not systematic and no comparison of the corresponding hardware implementations was made.

\emph{Contribution:} In this paper, we compare the error-correction performance and hardware efficiency of polar decoders---for the three main decoding algorithms---with those of decoders for the LDPC codes used in the IEEE~802.11ad (WiGig)~\cite{IEEE802.11ad}, IEEE~802.11n (Wi-Fi)~\cite{IEEE802.11n}, and IEEE~802.3an ($10$ Gb/s Ethernet)~\cite{IEEE802.3an} standards, as well as those of decoders for the Turbo code used in the 3GPP~LTE~\cite{LTE} standard. 

\section{Comparison Methodology}
Most hardware implementations of polar decoders in the literature focused on either on SC, BP, or SCL decoding algorithms. Thus, our comparison is for polar decoders based on these algorithms. The comparison against LDPC and Turbo decoders has two aspects as we are interested in both error-correction capabilities and hardware efficiency.

For the error-correction performance comparison of the various polar decoders with that of LDPC and Turbo decoders, floating-point versions of all decoding algorithms are used. The quantization parameters of the hardware decoders are usually chosen so that the performance loss with respect to the floating-point implementation is negligible. Moreover, for all simulations the encoded codewords obtained from random data are modulated using binary phase-shift keying (BPSK) and are transmitted over an additive white Gaussian noise (AWGN) channel. For almost all decoders for polar and LDPC codes, (scaled or offset) \emph{min-sum} approximations are used for check node updates. The scaling and/or offset factors are given, whenever applicable. Our Turbo decoder uses the \emph{max-log} approximation. All polar codes are designed using the Monte Carlo based method proposed by Ar{\i}kan~\cite{Arikan2009}. In order to speed up our simulations of BP decoding for polar codes, we used the $\mathbf{G}$ matrix based early termination method of~\cite{Yuan2014}, which has negligible impact on the error-correction performance. For the CRC-aided SCL decoders, we use the following CRC polynomial $g_8(x)    = x^8 + x^5 + x^4 + x^3 + 1$. The hardware comparison is performed by selecting parameters for the polar decoders (e.g., blocklength, list size, number of iterations) that lead to an error-correction performance that is close to that of the competing LDPC or Turbo codes. Unfortunately, power results for polar decoders are scarce in the literature, making a useful power comparison with existing LDPC and Turbo decoders difficult. Thus, only area and decoding time complexity (which is the inverse of the decoding throughput) are considered for the comparison. We plot these metrics against each other on a double-logarithmic plot where the area and time complexity are on the vertical and horizontal axes, respectively. We note that hardware efficiency is defined as unit area per decoded bit and is measured in mm$^2$/bits/s. Thus, on the aforementioned double-logarithmic plots, lines with a slope of $-1$ correspond to iso-hardware efficiency lines. 

In order to scale the area of all decoders appropriately, the following assumptions are made. First, all synthesis results are scaled to a $90$~nm CMOS technology using standard Dennard scaling laws~\cite{Dennard1974}, so that the area scales as $s^2$ and the operating frequency scales as $1/s$, where $s$ is the technology feature size. Moreover, the area of the SC and SCL decoders scales linearly with the blocklength, and the area of the BP decoders scales as $N \log N$. The area of the SCL decoders scales linearly with the list size. The decoding latency of the BP decoders scales linearly with the maximum number of iterations. As it is very difficult to predict the frequency scaling with respect to the blocklength and list size parameters, we only use technology scaling for the operating frequency as already explained. 

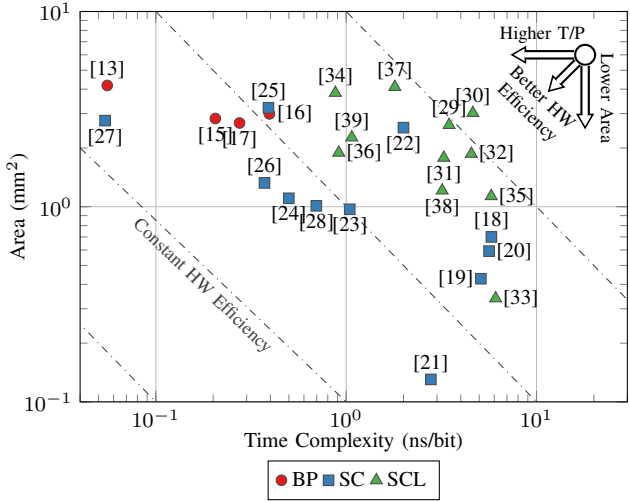
\begin{figure}
	\centering
	\begin{tikzpicture}
	\footnotesize
	\pgfplotsset{grid style={solid}}

	\begin{loglogaxis}[
		width = \figurewidth\columnwidth,
		height = 0.9*\figureheight\columnwidth,
		ylabel style={yshift=-0.4cm},
		xlabel = {Time Complexity (ns/bit)},
		ylabel = {Area (mm$^2$)},
    xlabel style={yshift=0.8em},%
		xmin = 4e-2, xmax = 3e1,
		ymin = 1e-1, ymax = 1e1,
		grid = major,
		legend style={at={(0.5,-0.15)},anchor=north},
		legend cell align=left,
		legend columns=4,
	]		

		\addplot[draw=black!70!white,fill=Set1-7-1, only marks, mark=*, 
				visualization depends on={value \thisrow{anchor}\as\myanchor},
				nodes near coords, 
				every node near coord/.append style={color=black}, 
			  every node near coord/.append style={anchor=\myanchor},
				point meta=explicit symbolic] table[meta=label] {images/data/hardware/BPHardwareEff.dat};
		\addlegendentry{BP}

		\addplot[draw=black!70!white,fill=Set1-7-2, only marks, mark=square*, 
				visualization depends on={value \thisrow{anchor}\as\myanchor},
				nodes near coords, 
				every node near coord/.append style={color=black}, 
			  every node near coord/.append style={anchor=\myanchor},
				point meta=explicit symbolic] table[meta=label] {images/data/hardware/SCHardwareEff.dat};
		\addlegendentry{SC}

		\addplot[draw=black!70!white,fill=Set1-7-3, only marks, mark=triangle*, mark options={scale=1.3, solid},
				visualization depends on={value \thisrow{anchor}\as\myanchor},
				nodes near coords, 
				every node near coord/.append style={color=black}, 
			  every node near coord/.append style={anchor=\myanchor},
				point meta=explicit symbolic] table[meta=label] {images/data/hardware/SCLHardwareL4.dat};
		\addlegendentry{SCL}

	  \addplot[mark=none, black, dashdotted, opacity=0.75] coordinates {(1e-2,1e-1) (1e-1,1e-2)};
	  \addplot[mark=none, black, dashdotted, opacity=0.75] coordinates {(1e-2,1e0) (1e0,1e-2)};
	  \addplot[mark=none, black, dashdotted, opacity=0.75] coordinates {(4e-2,2e0) (1e0,1e-1)} node [below, pos=0.5,rotate=-44] {\scriptsize Constant HW Efficiency};
	  \addplot[mark=none, black, dashdotted, opacity=0.75] coordinates {(1e-2,1e2) (1e2,1e-2)};
	  \addplot[mark=none, black, dashdotted, opacity=0.75] coordinates {(1e-1,1e2) (1e2,1e-1)};

		\node[draw,circle,thick] at (axis cs:18,6) (a) {};
		\node[inner sep=0,minimum size=0,left of=a] (k1) {};
		\node[inner sep=0,minimum size=0,below of=a] (k2) {};
		\node[inner sep=0,minimum size=0,below left of=a,yshift=0.03cm,xshift=0.03cm,rotate=-44,font=\scriptsize,text width=1.6cm,align=center] (k3) {Better HW Efficiency};
		\draw[vecArrow] (a) -- (k1) node[draw=none,fill=none,font=\scriptsize,midway,above,yshift=0.05cm] {Higher T/P};
		\draw[vecArrow] (a) -- (k2) node[draw=none,fill=none,font=\scriptsize,midway,above,rotate=-90,yshift=0.1cm] {Lower Area};
		\draw[vecArrow] (a) -- (k3) node[draw=none,fill=none,font=\scriptsize,above,yshift=0.05cm] {};

		\end{loglogaxis}

\end{tikzpicture}%
	\caption{Time complexity Vs area for various $N=1024$ polar decoders. SCL decoder implementations are given for~${L=4}$.}
	\label{fig:polaroverview}
\end{figure}

\section{Comparison of Polar Codes with LDPC and Turbo Codes}\label{sec:comparison}
Figure~\ref{fig:polaroverview} provides a summary of the area and time complexity of ASIC implementations of SC, BP, and SCL polar decoders used as reference for the comparison with the LDPC and Turbo codes. All decoders are scaled to $N=1024$, and the SCL decoders are scaled to $L=4$. 

We observe that BP decoders generally provide very high throughputs, although they are matched by some of the most recent fast-SSC-based SC decoders. We note that the fast-SSC decoder of \cite{Giard2016b} is specialized for a small set of polar codes and that BP decoding provides soft output values, which are required for iterative receivers. Moreover, the BP decoders also generally have the highest area requirements of all decoders. SCL decoders generally have the lowest throughput of all decoders, as well as higher area requirements than SC decoders and similar area requirements to BP decoders. However, SCL decoders provide significantly improved error-correction performance with respect to both SC and BP decoding.

\subsection{Polar Codes vs. IEEE 802.11ad LDPC Codes}
The IEEE 802.11ad standard~\cite{IEEE802.11ad} uses QC-LDPC codes with a blocklength of $N = 672$ and code rates $R \in \left\{\frac{1}{2}, \frac{5}{8}, \frac{3}{4}, \frac{13}{16}\right\}$. We simulated the performance of this LDPC code using a layered offset min-sum decoding algorithm with a maximum of $I = 5$ iterations and an offset of $\beta = 0.2$, which are numbers commonly found in the literature. A comparison for the lowest and highest rates $\left(R \in \left\{\frac{1}{2},\frac{13}{16}\right\}\right)$ found in the IEEE 802.11ad standard is provided.

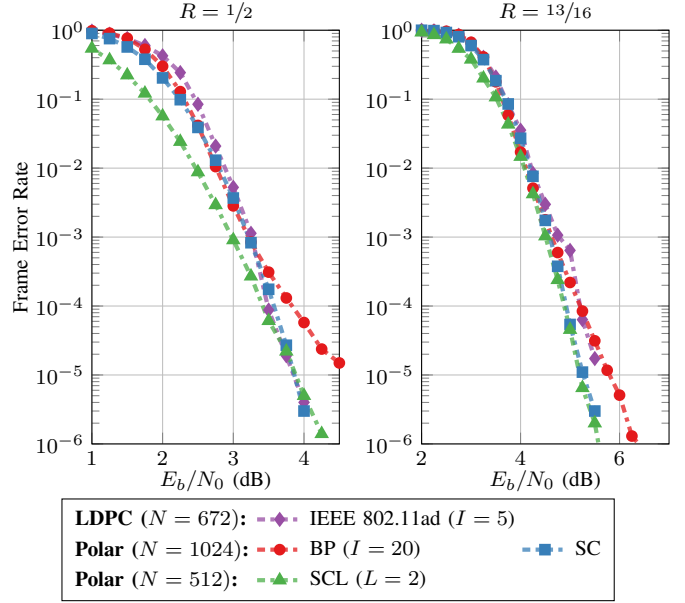
\begin{figure}
	\centering
	\begin{tikzpicture}
	\footnotesize
	\pgfplotsset{grid style={solid}}

	\begin{semilogyaxis}[
		width = \doublefigurewidth\columnwidth,
		height = \ferfigureheight\columnwidth,
		title = {$R = \sfrac{1}{2}$},
		title style={yshift=-0.8em},
		xlabel = {$E_b/N_0$ (dB)},
		xlabel style={yshift=1.0em},%
		ylabel = {Frame Error Rate},
		ylabel style={yshift=-1.0em},
		xmin = 1, xmax = 4.5,
		ymin = 1e-6, ymax = 1e0,
		grid = major,
		legend style={at={(0.45,-0.22)},anchor=north,font=\footnotesize},
		legend cell align=left,
		legend columns=3,
		legend to name=802.11advspolarEquiFER,
	]		

		\addlegendimage{empty legend}
		\addlegendentry{\bf LDPC ($N=672)$:}
		\addplot[color=Set1-4-4, opacity=0.75, ultra thick, dashdotted, mark=diamond*, mark options={scale=0.8, solid, opacity=1}] table {images/data/802.11ad/RES_IEEE11adD5_AWGN_Layered_R0.50_Z42_A_OMS_0_none_I5_B0.200_fp.dat};
		\addlegendentry{IEEE 802.11ad ($I=5$)}
		\addlegendimage{empty legend}
		\addlegendentry{}

		\addlegendimage{empty legend}
		\addlegendentry{\bf Polar ($N=1024)$:}
		\addplot[color=Set1-4-1, opacity=0.75, ultra thick, dashdotted, mark=*, mark options={scale=0.7, solid, opacity=1}] table {images/data/polar/BP/AWGN_N1024_SNR1.00_R0.50_BP_Iter20_Quant0_.dat};
		\addlegendentry{BP ($I=20$)}
		\addplot[color=Set1-4-2, opacity=0.75, ultra thick, dashdotted, mark=square*, mark options={scale=0.7, solid, opacity=1}] table {images/data/polar/SC/AWGN_N1024_SNR1.00_R0.50_SC_Quant0_.dat};
		\addlegendentry{SC}

		\addlegendimage{empty legend}
		\addlegendentry{\bf Polar ($N=512)$:}
		\addplot[color=Set1-4-3, opacity=0.75, ultra thick, dashdotted, mark=triangle*, mark options={scale=0.8, solid, opacity=1}] table {images/data/polar/SCL/AWGN_N512_SNR1.00_R0.50_SCListCRC_CRC8_L2_Quant0_.dat};
		\addlegendentry{SCL ($L=2$)}

		\end{semilogyaxis}

\end{tikzpicture}%
\begin{tikzpicture}
	\footnotesize
	\pgfplotsset{grid style={solid}}

	\begin{semilogyaxis}[
		width = \doublefigurewidth\columnwidth,
		height = \ferfigureheight\columnwidth,
		title = {$R = \sfrac{13}{16}$},
    title style={yshift=-0.8em},
		xlabel = {$E_b/N_0$ (dB)},
    xlabel style={yshift=1.0em},%
    ylabel style={yshift=-1.0em},
		xmin = 2, xmax = 7,
		ymin = 1e-6, ymax = 1e0,
		grid = major,
	]		

		\addplot[color=Set1-4-4, opacity=0.75, ultra thick, dashdotted, mark=diamond*, mark options={scale=0.8, solid, opacity=1}] table {images/data/802.11ad/RES_IEEE11adD5_AWGN_Layered_R0.81_Z42_A_OMS_0_none_I5_B0.200_fp.dat};
		\addplot[color=Set1-4-1, opacity=0.75, ultra thick, dashdotted, mark=*, mark options={scale=0.7, solid, opacity=1}] table {images/data/polar/BP/AWGN_N1024_SNR4.00_R0.81_BP_Iter20_Quant0_.dat};
		\addplot[color=Set1-4-2, opacity=0.75, ultra thick, dashdotted, mark=square*, mark options={scale=0.7, solid, opacity=1}] table {images/data/polar/SC/AWGN_N1024_SNR4.00_R0.81_SC_Quant0_.dat};
		\addplot[color=Set1-4-3, opacity=0.75, ultra thick, dashdotted, mark=triangle*, mark options={scale=0.8, solid, opacity=1}] table {images/data/polar/SCL/AWGN_N512_SNR4.00_R0.81_SCListCRC_CRC8_L2_Quant0_.dat};

		\end{semilogyaxis}

\end{tikzpicture}%
\\%
\ref{802.11advspolarEquiFER}%
	\caption{Performance of the LDPC code of the IEEE 802.11ad standard compared to polar codes.}
	\label{fig:802.11adequiFER}
\end{figure}%
\begin{figure}
	\centering
	\begin{tikzpicture}
	\footnotesize
	\pgfplotsset{grid style={solid}}

	\begin{loglogaxis}[
		width = \figurewidth\columnwidth,
		height = \figureheight\columnwidth,
		ylabel style={yshift=-0.4cm},
		xlabel = {Time Complexity (ns/bit)},
		ylabel = {Area (mm$^2$)},
		xlabel style={yshift=0.8em},%
		xmin = 5e-2, xmax = 1e1,
		ymin = 1e-1, ymax = 1e1,
		grid = major,
		legend style={at={(.5,-0.125)},anchor=north},
		legend cell align=left,
		legend columns=4,
	]		

		\addplot[draw=black!70!white,fill=Set1-7-4, only marks, mark=diamond*, , mark options={scale=1.3, solid},
				visualization depends on={value \thisrow{anchor}\as\myanchor},
				nodes near coords, 
				every node near coord/.append style={color=black}, 
			  every node near coord/.append style={anchor=\myanchor},
				point meta=explicit symbolic] table[meta=label] {images/data/hardware/ADHardwareEff.dat};
		\addlegendentry{IEEE 802.11ad}

		\addplot[draw=black!70!white,fill=Set1-7-1, only marks, mark=*] table[x expr=\thisrowno{0}, y expr=\thisrowno{1}] {images/data/hardware/BPHardwareEff.dat};
		\addlegendentry{BP}

		\addplot[draw=black!70!white,fill=Set1-7-2, only marks, mark=square*] table[x expr=\thisrowno{0}, y expr=\thisrowno{1}] {images/data/hardware/SCHardwareEff.dat};
		\addlegendentry{SC}

		\addplot[draw=black!70!white,fill=Set1-7-3, only marks, mark=triangle*, mark options={scale=1.3, solid}] table[x expr=\thisrowno{0}, y expr=\thisrowno{1}/2] {images/data/hardware/SCLHardwareL2.dat};
		\addlegendentry{SCL}

	  \addplot[mark=none, black, dashdotted, opacity=0.75] coordinates {(1e-2,1e0) (1e-1,1e-1)};
	  \addplot[mark=none, black, dashdotted, opacity=0.75] coordinates {(1e-2,1e1) (1e0,1e-1)} node [below, pos=0.7,rotate=-43] {\scriptsize Constant HW Efficiency};
	  \addplot[mark=none, black, dashdotted, opacity=0.75] coordinates {(1e-1,1e1) (1e1,1e-1)};
	  \addplot[mark=none, black, dashdotted, opacity=0.75] coordinates {(1e0,1e1) (1e1,1e0)};
	
		\node[draw,circle,thick] at (axis cs:6.8,6.2) (a) {};
		\node[inner sep=0,minimum size=0,left of=a] (k1) {};
		\node[inner sep=0,minimum size=0,below of=a] (k2) {};
		\node[inner sep=0,minimum size=0,below left of=a,yshift=0.03cm,xshift=0.03cm,rotate=-43,font=\scriptsize,text width=1.6cm,align=center] (k3) {Better HW Efficiency};
		\draw[vecArrow] (a) -- (k1) node[draw=none,fill=none,font=\scriptsize,midway,above,yshift=0.05cm] {Higher T/P};
		\draw[vecArrow] (a) -- (k2) node[draw=none,fill=none,font=\scriptsize,midway,above,rotate=-90,yshift=0.1cm] {Lower Area};
		\draw[vecArrow] (a) -- (k3) node[draw=none,fill=none,font=\scriptsize,above,yshift=0.05cm] {};

		\end{loglogaxis}

\end{tikzpicture}%
	\caption{Hardware efficiency of IEEE 802.11ad LDPC decoders against that of polar decoders.}
	\label{fig:802.11adequiFERhardware}
\end{figure}
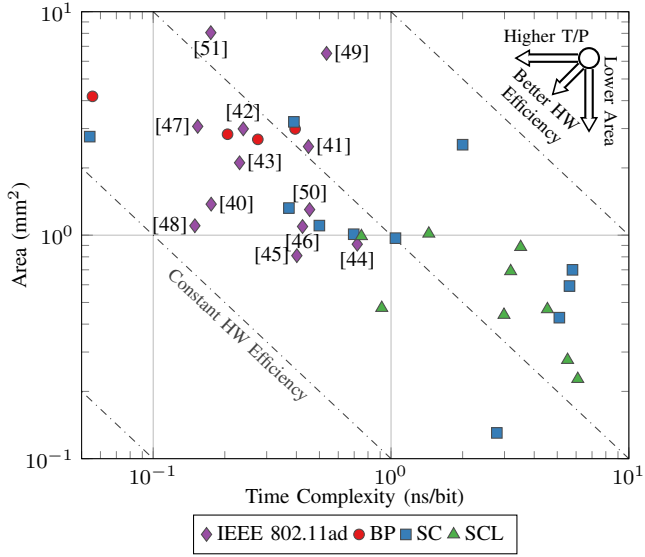%
Figure~\ref{fig:802.11adequiFER} shows that SC and BP decoding of a $N=1024$ polar code performs very similarly to the LDPC codes of the IEEE 802.11ad standard. Moreover, SCL decoding of a $N=512$ polar code with $L=2$ and an 8-bit CRC is sufficient to match the error-correction performance of the LDPC code. We note that both the $N=512$ codes used for SCL decoding and the $N=1024$ codes used for SC and BP decoding were designed for an SNR of $1$~dB and $4$~dB for $R = \frac{1}{2}$ and $R = \frac{13}{16}$, respectively.

From Figure~\ref{fig:802.11adequiFERhardware}, it can be seen that all BP as well as the best SC polar decoders compete well in terms of area, throughput, and hardware efficiency against LDPC decoders. While the hardware efficiency of SCL decoders is similar to IEEE 802.11ad LDPC decoders due to their lower area requirements, most SCL decoders have lower throughput.%

\subsection{Polar Codes vs. IEEE 802.11n LDPC Codes}
The IEEE 802.11n standard~\cite{IEEE802.11n} uses QC-LDPC codes with blocklengths of $N \in \{648,1296,1944\}$ and code rates $R \in \left\{\frac{1}{2}, \frac{2}{3}, \frac{3}{4}, \frac{5}{6}\right\}$. We simulated the performance of this LDPC code using a layered offset min-sum decoding algorithm with a maximum of $I = 12$ iterations and an offset of $\beta = 0.5$, which are  numbers commonly found in the literature. We provide a comparison for $N = 1944$ and for the lowest rate $\left(R = \frac{1}{2}\right)$ and the highest rate $\left(R=\frac{5}{6}\right)$ found in the IEEE 802.11n standard. 

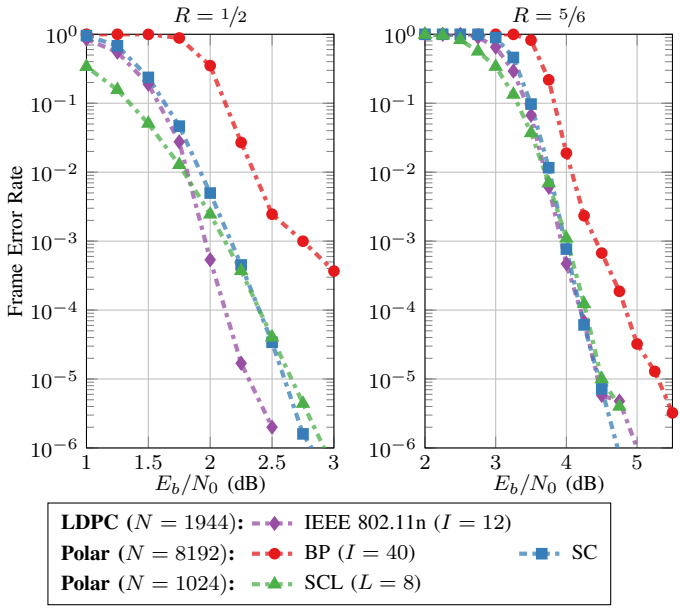
\begin{figure}
	\centering
	\begin{tikzpicture}
	\footnotesize
	\pgfplotsset{grid style={solid}}

	\begin{semilogyaxis}[
		width = \doublefigurewidth\columnwidth,
		height = \ferfigureheight\columnwidth,
		title = {$R = \sfrac{1}{2}$},
		xlabel = {$E_b/N_0$ (dB)},
		ylabel = {Frame Error Rate},
    title style={yshift=-0.8em},
    xlabel style={yshift=1.0em},%
    ylabel style={yshift=-1.0em},
		xmin = 1, xmax = 3,
		ymin = 1e-6, ymax = 1e0,
		grid = major,
		legend style={at={(0.45,-0.22)},anchor=north,font=\footnotesize},
		legend cell align=left,
		legend columns=3,
		legend to name=802.11nEquiFERvspolar,
	]		

		\addlegendimage{empty legend}
		\addlegendentry{\bf LDPC ($N=1944)$:}
		\addplot[Set1-4-4, opacity=0.75, ultra thick, dashdotted, mark=diamond*, mark options={scale=0.8, solid, opacity=1}] table {images/data/802.11n/RES_IEEE11nD7_AWGN_Layered_R0.50_Z81_A_OMS_0_none_I12_B0.500_fp.dat};
		\addlegendentry{IEEE 802.11n ($I=12$)}
		\addlegendimage{empty legend}
		\addlegendentry{}

		\addlegendimage{empty legend}
		\addlegendentry{\bf Polar ($N=8192)$:}
		\addplot[Set1-4-1, opacity=0.75, ultra thick, dashdotted, mark=*, mark options={scale=0.7, solid, opacity=1}] table {images/data/polar/BP/AWGN_N8192_SNR0.00_R0.50_BP_Iter40_Quant0_.dat};
		\addlegendentry{BP ($I=40$)}
		\addplot[Set1-4-2, opacity=0.75, ultra thick, dashdotted, mark=square*, mark options={scale=0.7, solid, opacity=1}] table {images/data/polar/SC/AWGN_N8192_SNR-1.00_R0.50_SC_Quant0_.dat};
		\addlegendentry{SC}
		\addlegendimage{empty legend}
		\addlegendentry{\bf Polar ($N=1024)$:}
		\addplot[Set1-4-3, opacity=0.75, ultra thick, dashdotted, mark=triangle*, mark options={scale=0.8, solid, opacity=1}] table {images/data/polar/SCL/AWGN_N1024_SNR0.00_R0.50_SCListCRC_CRC8_L8_Quant0_.dat};
		\addlegendentry{SCL ($L=8$)}

		\end{semilogyaxis}

\end{tikzpicture}%
\begin{tikzpicture}
	\footnotesize
	\pgfplotsset{grid style={solid}}

	\begin{semilogyaxis}[
		width = \doublefigurewidth\columnwidth,
		height = \ferfigureheight\columnwidth,
		title = {$R = \sfrac{5}{6}$},
		xlabel = {$E_b/N_0$ (dB)},
    title style={yshift=-0.8em},
    xlabel style={yshift=1.0em},%
    ylabel style={yshift=-1.0em},
		xmin = 2, xmax = 5.5,
		ymin = 1e-6, ymax = 1e0,
		grid = major,
	]		

		\addplot[Set1-4-4, opacity=0.75, ultra thick, dashdotted, mark=diamond*, mark options={scale=0.8, solid, opacity=1}] table {images/data/802.11n/RES_IEEE11nD7_AWGN_Layered_R0.83_Z81_A_OMS_0_none_I12_B0.500_fp.dat};
		\addplot[Set1-4-1, opacity=0.75, ultra thick, dashdotted, mark=*, mark options={scale=0.7, solid, opacity=1}] table {images/data/polar/BP/AWGN_N8192_SNR4.00_R0.83_BP_Iter40_Quant0_.dat};
		\addplot[Set1-4-2, opacity=0.75, ultra thick, dashdotted, mark=square*, mark options={scale=0.7, solid, opacity=1}] table {images/data/polar/SC/AWGN_N8192_SNR3.00_R0.83_SC_Quant0_.dat};
		\addplot[Set1-4-3, opacity=0.75, ultra thick, dashdotted, mark=triangle*, mark options={scale=0.8, solid, opacity=1}] table {images/data/polar/SCL/AWGN_N1024_SNR4.00_R0.83_SCListCRC_CRC8_L8_Quant0_.dat};

		\end{semilogyaxis}

\end{tikzpicture}%
\\
\ref{802.11nEquiFERvspolar}
	\caption{Performance of the LDPC code of IEEE 802.11n standard compared to polar codes.}
	\label{fig:802.11nEquiFER}
\end{figure}%
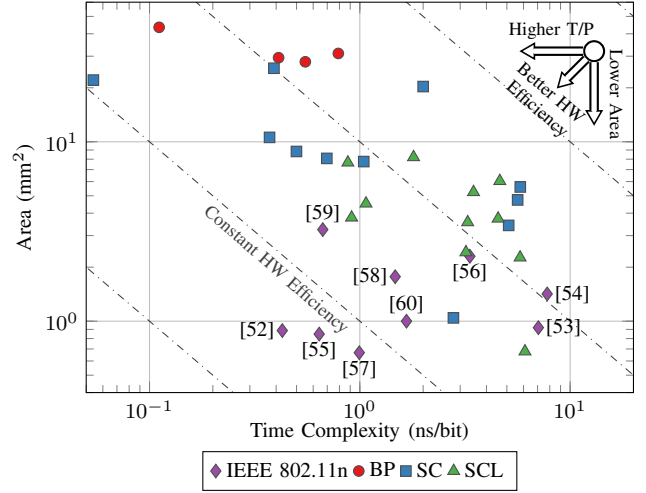
\begin{figure}
	\centering
	\begin{tikzpicture}
	\footnotesize
	\pgfplotsset{grid style={solid}}

	\begin{loglogaxis}[
		width = \figurewidth\columnwidth,
		height = 0.9*\figureheight\columnwidth,
		ylabel style={yshift=-0.4cm},
		xlabel = {Time Complexity (ns/bit)},
		ylabel = {Area (mm$^2$)},
    xlabel style={yshift=0.8em},%
		xmin = 5e-2, xmax = 2e1,
		ymin = 4e-1, ymax = 6e1,
		grid = major,
		legend style={at={(0.5,-0.15)},anchor=north},
		legend cell align=left,
		legend columns=4,
	]		

		\addplot[draw=black!70!white,fill=Set1-7-4, only marks, mark=diamond*, , mark options={scale=1.3, solid},
				visualization depends on={value \thisrow{anchor}\as\myanchor},
				nodes near coords, 
				every node near coord/.append style={color=black}, 
			  every node near coord/.append style={anchor=\myanchor},
				point meta=explicit symbolic] table[meta=label] {images/data/hardware/NHardwareEff.dat};
		\addlegendentry{IEEE 802.11n}

		\addplot[draw=black!70!white,fill=Set1-7-1, only marks, mark=*] table[x expr=\thisrowno{0}*2, y expr=\thisrowno{1}*(8*(13/10)] {images/data/hardware/BPHardwareEff.dat};
		\addlegendentry{BP}

		\addplot[draw=black!70!white,fill=Set1-7-2, only marks, mark=square*] table[x expr=\thisrowno{0}, y expr=\thisrowno{1}*8] {images/data/hardware/SCHardwareEff.dat};
		\addlegendentry{SC}

		\addplot[draw=black!70!white,fill=Set1-7-3, only marks, mark=triangle*, mark options={scale=1.3, solid}] table[x expr=\thisrowno{0}, y expr=\thisrowno{1}*2] {images/data/hardware/SCLHardwareL4.dat};
		\addlegendentry{SCL}

	  \addplot[mark=none, black, dashdotted, opacity=0.75] coordinates {(1e-2,1e1) (1e0,1e-1)};
	  \addplot[mark=none, black, dashdotted, opacity=0.75] coordinates {(1e-2,1e2) (1e1,1e-1)} node [below, pos=0.55,rotate=-41] {\scriptsize Constant HW Efficiency};
	  \addplot[mark=none, black, dashdotted, opacity=0.75] coordinates {(1e-1,1e2) (1e2,1e-1)};
	  \addplot[mark=none, black, dashdotted, opacity=0.75] coordinates {(1e0,1e2) (1e2,1e0)};
	  \addplot[mark=none, black, dashdotted, opacity=0.75] coordinates {(1e1,1e2) (1e2,1e1)};
	
		\node[draw,circle,thick] at (axis cs:13,32) (a) {};
		\node[inner sep=0,minimum size=0,left of=a] (k1) {};
		\node[inner sep=0,minimum size=0,below of=a] (k2) {};
		\node[inner sep=0,minimum size=0,below left of=a,yshift=0.03cm,xshift=0.03cm,rotate=-41,font=\scriptsize,text width=1.6cm,align=center] (k3) {Better HW Efficiency};
		\draw[vecArrow] (a) -- (k1) node[draw=none,fill=none,font=\scriptsize,midway,above,yshift=0.05cm] {Higher T/P};
		\draw[vecArrow] (a) -- (k2) node[draw=none,fill=none,font=\scriptsize,midway,above,rotate=-90,yshift=0.1cm] {Lower Area};
		\draw[vecArrow] (a) -- (k3) node[draw=none,fill=none,font=\scriptsize,above,yshift=0.05cm] {};

		\end{loglogaxis}

\end{tikzpicture}%
	\caption{Hardware efficiency of IEEE 802.11n LDPC decoders against that of polar decoders.}
	\label{fig:802.11nhardwareEquiFER}
\end{figure}%

In Figure~\ref{fig:802.11nEquiFER}, we observe that a polar code with $N=8192$ under SC decoding has a small loss of $0.5$~dB with respect to the IEEE 802.11n LDPC code with $N=1944$ at a FER of $10^{-5}$ for $R=\frac{1}{2}$, while the error-correction performance for $R=\frac{5}{6}$ is very similar. Moreover, a polar code with $N=1024$ under SCL decoding with $L=8$ and an $8$-bit CRC has practically identical performance with the aforementioned polar code with $N=8192$ under SC decoding for both $R=\frac{1}{2}$ and $R=\frac{5}{6}$. Unfortunately, the polar code with $N=8192$ under BP decoding cannot reach the performance of the IEEE 802.11n LDPC code, even when a maximum of $I=40$ iterations are performed. We note that the polar codes with $N=8192$ used for SC and BP decoding were designed for an SNR of $-1$~dB and $3$~dB for $R=\frac{1}{2}$ and $\frac{5}{6}$, respectively, while the polar codes with $N=1024$ used for SCL decoding with $L=8$ were designed for an SNR of $0$~dB and $4$~dB for $R=\frac{1}{2}$ and $\frac{5}{6}$, respectively.

In Figure~\ref{fig:802.11nhardwareEquiFER}, we observe that, on average, the SCL decoders have the highest hardware efficiency out of the polar decoders. Both the SC and the BP decoders have significantly higher area requirements when trying to match the FER performance of the IEEE 802.11n LDPC codes. Finally, we observe that, on average, the IEEE 802.11n LDPC decoders have a slightly higher hardware efficiency than the polar decoders.%

\subsection{Polar Codes vs. IEEE 802.3an LDPC Codes}
The IEEE 802.3an standard~\cite{IEEE802.3an} uses a $(6,32)$-regular LDPC code with blocklength $N=2048$ and code design rate $R = \frac{13}{16}$. In our simulations, the LDPC code is decoded using a flooding sum-product decoder with $I=8$ maximum decoding iterations, which is a number that is commonly found in the literature (we note that $4$-$5$ layered iterations provide similar error-correction performance to $8$-$10$ flooding iterations).

\begin{figure}
	\centering
	\begin{tikzpicture}
	\footnotesize
	\pgfplotsset{grid style={solid}}

	\begin{semilogyaxis}[
		width = \figurewidth\columnwidth,
		height = \ferfigureheight\columnwidth,
		title = {$R = \sfrac{13}{16}$},
		xlabel = {$E_b/N_0$ (dB)},
		ylabel = {Frame Error Rate},
    title style={yshift=-0.8em},
    xlabel style={yshift=1.0em},%
    ylabel style={yshift=-1.0em},
		xmin = 2.5, xmax = 5,
		ymin = 1e-6, ymax = 1e0,
		grid = major,
		legend style={at={(0.4,-0.17)},anchor=north,font=\footnotesize},
		legend cell align=left,
		legend columns=3,
	]		

		\addlegendimage{legend image with text=\textbf{LDPC}}
		\addlegendentry{($N=2048)$:}
		\addplot[Set1-4-4, opacity=0.75, ultra thick, dashdotted, mark=diamond*, mark options={scale=0.8, solid, solid, opacity=1}] table {images/data/802.3an/RES_N2048_R0.81_v6c32-reg_maxIter08_SumProd_float_channels1E+04.dat};
		\addlegendentry{IEEE 802.3an ($I=8)$}
		\addlegendimage{empty legend}
		\addlegendentry{}

		\addlegendimage{legend image with text=\textbf{Polar}}
		\addlegendentry{($N=4096)$:}
		\addplot[Set1-4-1, opacity=0.75, ultra thick, dashdotted, mark=*, mark options={scale=0.7, solid, solid, opacity=1}] table {images/data/polar/BP/AWGN_N4096_SNR3.00_R0.81_BP_Iter40_Quant0_.dat};
		\addlegendentry{BP ($I=40$)}
		\addplot[Set1-4-2, opacity=0.75, ultra thick, dashdotted, mark=square*, mark options={scale=0.7, solid, solid, opacity=1}] table {images/data/polar/SC/AWGN_N4096_SNR3.00_R0.81_SC_Quant0_.dat};
		\addlegendentry{SC}
		\addlegendimage{legend image with text=\textbf{Polar}}
		\addlegendentry{($N=1024)$:}
		\addplot[Set1-4-3, opacity=0.75, ultra thick, dashdotted, mark=triangle*, mark options={scale=0.8, solid, solid, opacity=1}] table {images/data/polar/SCL/AWGN_N1024_SNR4.00_R0.81_SCListCRC_CRC8_L4_Quant0_.dat};
		\addlegendentry{SCL ($L=4$)}

		\end{semilogyaxis}

\end{tikzpicture}%
	\caption{Performance of the LDPC code of the IEEE 802.3an standard compared to polar codes.}
	\label{fig:802.3anEquiFER}
\end{figure}%
\begin{figure}
	\centering
	\begin{tikzpicture}
	\footnotesize
	\pgfplotsset{grid style={solid}}

	\begin{loglogaxis}[
		width = \figurewidth\columnwidth,
		height = 0.9*\figureheight\columnwidth,
		ylabel style={yshift=-0.4cm},
		xlabel = {Time Complexity (ns/bit)},
		ylabel = {Area (mm$^2$)},
    xlabel style={yshift=0.8em},%
		xmin = 2e-2, xmax = 2.5e1,
		ymin = 2e-1, ymax = 3e1,
		grid = major,
		legend style={at={(0.5,-0.15)},anchor=north},
		legend cell align=left,
		legend columns=4,
	]		

		\addplot[draw=black!70!white,fill=Set1-7-4, only marks, mark=diamond*, , mark options={scale=1.3, solid},
				visualization depends on={value \thisrow{anchor}\as\myanchor},
				nodes near coords, 
				every node near coord/.append style={color=black}, 
			  every node near coord/.append style={anchor=\myanchor},
				point meta=explicit symbolic] table[meta=label] {images/data/hardware/GE10HardwareEff.dat};
		\addlegendentry{IEEE 802.3an}

		\addplot[draw=black!70!white,fill=Set1-7-1, only marks, mark=*] table[x expr=\thisrowno{0}*2, y expr=\thisrowno{1}*(4*12/10)] {images/data/hardware/BPHardwareEff.dat};
		\addlegendentry{BP}

		\addplot[draw=black!70!white,fill=Set1-7-2, only marks, mark=square*] table[x expr=\thisrowno{0}, y expr=\thisrowno{1}*4] {images/data/hardware/SCHardwareEff.dat};
		\addlegendentry{SC}

		\addplot[draw=black!70!white,fill=Set1-7-3, only marks, mark=triangle*, mark options={scale=1.3, solid}] table[x expr=\thisrowno{0}, y expr=\thisrowno{1}] {images/data/hardware/SCLHardwareL4.dat};
		\addlegendentry{SCL}

	  \addplot[mark=none, black, dashdotted, opacity=0.75] coordinates {(1e-2,1e0) (1e-1,1e-1)};
	  \addplot[mark=none, black, dashdotted, opacity=0.75] coordinates {(1e-2,1e1) (1e0,1e-1)} node [below, pos=0.5,rotate=-44.5] {\scriptsize Constant HW Efficiency};
	  \addplot[mark=none, black, dashdotted, opacity=0.75] coordinates {(1e-2,1e2) (1e1,1e-1)};
	  \addplot[mark=none, black, dashdotted, opacity=0.75] coordinates {(1e-1,1e2) (1e2,1e-1)};
	  \addplot[mark=none, black, dashdotted, opacity=0.75] coordinates {(1e0,1e2) (1e2,1e0)};
	  \addplot[mark=none, black, dashdotted, opacity=0.75] coordinates {(1e1,1e2) (1e2,1e1)};

		\node[draw,circle,thick] at (axis cs:15,16) (a) {};
		\node[inner sep=0,minimum size=0,left of=a] (k1) {};
		\node[inner sep=0,minimum size=0,below of=a] (k2) {};
		\node[inner sep=0,minimum size=0,below left of=a,yshift=0.03cm,xshift=0.03cm,rotate=-44.5,font=\scriptsize,text width=1.6cm,align=center] (k3) {Better HW Efficiency};
		\draw[vecArrow] (a) -- (k1) node[draw=none,fill=none,font=\scriptsize,midway,above,yshift=0.05cm] {Higher T/P};
		\draw[vecArrow] (a) -- (k2) node[draw=none,fill=none,font=\scriptsize,midway,above,rotate=-90,yshift=0.1cm] {Lower Area};
		\draw[vecArrow] (a) -- (k3) node[draw=none,fill=none,font=\scriptsize,above,yshift=0.05cm] {};
	
		\end{loglogaxis}

\end{tikzpicture}%
	\caption{Hardware efficiency of IEEE 802.3an LDPC decoders against that of polar decoders.}
	\label{fig:802.3anHardwareEquiFER}
\end{figure}%

SCL decoding with $N=1024$, $L=4$, and an $8$-bit CRC already performs better than the IEEE 802.3an LDPC code down to a FER of $10^{-6}$. In Figure~\ref{fig:802.3anEquiFER}, we observe that a polar code with $N=4096$ under SC decoding has better error-correction performance than the IEEE 802.3an LDPC code down to a FER of $10^{-6}$. BP decoding with $I=40$ for the same polar code, however, has a small loss of $0.5$~dB with respect to the IEEE 802.3an LDPC code at a FER of $10^{-5}$. We note, however, that the FER curve of the IEEE 802.3an LDPC code has a steeper slope and this code will thus perform better than polar codes at lower FERs. The polar code for $N=1024$ and $R = \frac{13}{16}$ used for SCL decoding was designed for an SNR of $4$~dB, while the polar code for $N=4096$ and $R = \frac{13}{16}$ used for SC and BP decoding was designed for an SNR of $3$~dB.

In Figure~\ref{fig:802.3anHardwareEquiFER}, we observe that, on average, the polar decoders have lower hardware efficiency than the IEEE 802.3an LDPC decoders. In terms of decoding throughput, only the BP decoders and a few SC decoders can approach the IEEE 802.3an LDPC decoders, albeit with slightly higher area requirements.%

\subsection{Polar Codes vs. 3GPP LTE Turbo Codes}
The 3GPP LTE standard~\cite{LTE} defines a baseline Turbo code with rate $R=\frac{1}{3}$ and information bit interleaver block sizes ranging from $K=40$ to $K=6144$ bits. Multiple code rates are supported, both higher and lower than $R=\frac{1}{3}$, which are obtained by puncturing and parity bit repetition, respectively. We simulated the performance of this Turbo code for the largest supported interleaver length $K=6144$ under max-log decoding with $I=6$ iterations, which is a number that is commonly found in the hardware implementation literature. We note that an interleaver length of $K=6144$ leads to a codeword blocklength $N=12288$ for rate $R = \frac{1}{2}$ and a codeword blocklength of $N=18432$ for rate $R=\frac{1}{3}$. We provide a comparison for $R=\frac{1}{3}$ and $R=\frac{1}{2}$.

\begin{figure}
	\centering
	\begin{tikzpicture}
	\footnotesize
	\pgfplotsset{grid style={solid}}

	\begin{semilogyaxis}[
		width = \doublefigurewidth\columnwidth,
		height = \ferfigureheight\columnwidth,
		title = {$R = \sfrac{1}{3}$},
		xlabel = {$E_b/N_0$ (dB)},
		ylabel = {Frame Error Rate},
    title style={yshift=-0.8em},
    xlabel style={yshift=1.0em},%
    ylabel style={yshift=-1.0em},
		xmin = 0, xmax = 3,
		ymin = 1e-6, ymax = 1e0,
		grid = major,
		legend style={at={(0.45,-0.22)},anchor=north,font=\footnotesize},
		legend cell align=left,
		legend columns=3,
		legend to name=LTEvspolarEquiFER,
	]		

		\addlegendimage{empty legend}
		\addlegendentry{\bf Turbo ($K=6144)$:}
		\addplot[Set1-4-4, ultra thick, dashdotted, mark=diamond*, mark options={scale=0.8, solid}] table {images/data/LTE/ERR_PCTC_LTE_V3_R13_6144b_I6.dat};
		\addlegendentry{3GPP LTE ($I=6)$}
		\addlegendimage{empty legend}
		\addlegendentry{}

		\addlegendimage{empty legend}
		\addlegendentry{\bf Polar ($N=16384)$:}
		\addplot[Set1-4-1, ultra thick, dashdotted, mark=*, mark options={scale=0.7, solid}] table {images/data/polar/BP/AWGN_N32768_SNR0.00_R0.33_BP_Iter30_Quant0_.dat};
		\addlegendentry{BP ($I=30$)}
		\addplot[Set1-4-2, ultra thick, dashdotted, mark=square*, mark options={scale=0.7, solid}] table {images/data/polar/SC/AWGN_N16384_SNR-3.00_R0.33_SC_Quant0_.dat};
		\addlegendentry{SC}
		\addlegendimage{empty legend}
		\addlegendentry{\bf Polar ($N=2048)$:}
		\addplot[Set1-4-3, ultra thick, dashdotted, mark=triangle*, mark options={scale=0.8, solid}] table {images/data/polar/SCL/AWGN_N2048_SNR-2.00_R0.33_SCListCRC_CRC8_L8_Quant0_.dat};
		\addlegendentry{SCL ($L=8$)}

		\end{semilogyaxis}

\end{tikzpicture}%
\begin{tikzpicture}
	\footnotesize
	\pgfplotsset{grid style={solid}}

	\begin{semilogyaxis}[
		width = \doublefigurewidth\columnwidth,
		height = \ferfigureheight\columnwidth,
		title = {$R = \sfrac{1}{2}$},
		xlabel = {$E_b/N_0$ (dB)},
    title style={yshift=-0.8em},
    xlabel style={yshift=1.0em},%
    ylabel style={yshift=-1.0em},
		xmin = 0.5, xmax = 3,
		ymin = 1e-6, ymax = 1e0,
		grid = major,
	]		

		\addplot[Set1-4-4, ultra thick, dashdotted, mark=diamond*, mark options={scale=0.8, solid}] table {images/data/LTE/ERR_PCTC_LTE_V3_R12_6144b_I6.dat};

		\addplot[Set1-4-1, ultra thick, dashdotted, mark=*, mark options={scale=0.7, solid}] table {images/data/polar/BP/AWGN_N32768_SNR1.00_R0.50_BP_Iter30_Quant0_.dat};
		\addplot[Set1-4-2, ultra thick, dashdotted, mark=square*, mark options={scale=0.7, solid}] table {images/data/polar/SC/AWGN_N16384_SNR-1.00_R0.50_SC_Quant0_.dat};
		\addplot[Set1-4-3, ultra thick, dashdotted, mark=triangle*, mark options={scale=0.8, solid}] table {images/data/polar/SCL/AWGN_N2048_SNR0.00_R0.50_SCListCRC_CRC8_L8_Quant0_.dat};

		\end{semilogyaxis}

\end{tikzpicture}%
\\
\ref{LTEvspolarEquiFER}%
	\caption{Performance of Turbo code of LTE standard compared to polar codes.}
	\label{fig:LTEequiFER}
\end{figure}%
\begin{figure}
	\centering
	\begin{tikzpicture}
	\footnotesize
	\pgfplotsset{grid style={solid}}

	\begin{loglogaxis}[
		width = \figurewidth\columnwidth,
		height = 0.9*\figureheight\columnwidth,
		ylabel style={yshift=-0.4cm},
		xlabel = {Time Complexity (ns/bit)},
		ylabel = {Area (mm$^2$)},
    xlabel style={yshift=0.8em},%
		xmin = 5e-2, xmax = 1.5e1,
		ymin = 1e-0, ymax = 1.5e2,
		grid = major,
		legend style={at={(0.5,-0.15)},anchor=north},
		legend cell align=left,
		legend columns=4,
	]		

		\addplot[draw=black!70!white,fill=Set1-7-4, only marks, mark=diamond*, , mark options={scale=1.3, solid},
				visualization depends on={value \thisrow{anchor}\as\myanchor},
				nodes near coords, 
				every node near coord/.append style={color=black}, 
			  every node near coord/.append style={anchor=\myanchor},
				point meta=explicit symbolic] table[meta=label] {images/data/hardware/LTEHardwareEff.dat};
		\addlegendentry{3GPP LTE}

		\addplot[draw=black!70!white,fill=Set1-7-1, only marks, mark=*] table[x expr=\thisrowno{0}*1.5, y expr=\thisrowno{1}*(16*14/10)] {images/data/hardware/BPHardwareEff.dat};
		\addlegendentry{BP}

		\addplot[draw=black!70!white,fill=Set1-7-2, only marks, mark=square*] table[x expr=\thisrowno{0}, y expr=\thisrowno{1}*16] {images/data/hardware/SCHardwareEff.dat};
		\addlegendentry{SC}

		\addplot[draw=black!70!white,fill=Set1-7-3, only marks, mark=triangle*, mark options={scale=1.3, solid}] table[x expr=\thisrowno{0}, y expr=\thisrowno{1}*4] {images/data/hardware/SCLHardwareL4.dat};
		\addlegendentry{SCL}

	  \addplot[mark=none, black, dashdotted, opacity=0.75] coordinates {(1e-2,1e2) (1e1,1e-1)} node [below, pos=0.45,rotate=-40] {\scriptsize Constant HW Efficiency};
	  \addplot[mark=none, black, dashdotted, opacity=0.75] coordinates {(1e-2,1e3) (1e2,1e-1)};
	  \addplot[mark=none, black, dashdotted, opacity=0.75] coordinates {(1e-1,1e3) (1e2,1e0)};
	  \addplot[mark=none, black, dashdotted, opacity=0.75] coordinates {(1e0,1e3) (1e2,1e1)};

		\node[draw,circle,thick] at (axis cs:10,80) (a) {};
		\node[inner sep=0,minimum size=0,left of=a] (k1) {};
		\node[inner sep=0,minimum size=0,below of=a] (k2) {};
		\node[inner sep=0,minimum size=0,below left of=a,yshift=0.03cm,xshift=0.03cm,rotate=-42,font=\scriptsize,text width=1.6cm,align=center] (k3) {Better HW Efficiency};
		\draw[vecArrow] (a) -- (k1) node[draw=none,fill=none,font=\scriptsize,midway,above,yshift=0.05cm] {Higher T/P};
		\draw[vecArrow] (a) -- (k2) node[draw=none,fill=none,font=\scriptsize,midway,above,rotate=-90,yshift=0.1cm] {Lower Area};
		\draw[vecArrow] (a) -- (k3) node[draw=none,fill=none,font=\scriptsize,above,yshift=0.05cm] {};
	
		\end{loglogaxis}

\end{tikzpicture}%
	\caption{Hardware efficiency of 3GPP LTE Turbo decoders against that of polar decoders.}
	\label{fig:LTEHardwareequiFER}
\end{figure}%
In Figure~\ref{fig:LTEequiFER}, we observe that a polar code with $N=16384$ under SC decoding has a small loss of $0.5$~dB with respect to the LTE Turbo code with $K=6144$ at a FER of $10^{-5}$ for both $R=\frac{1}{3}$ and $R=\frac{1}{2}$ and a polar code with $N=2048$ under SCL decoding with $L=8$ and an $8$-bit CRC has the same loss of $0.5$~dB with respect to the LTE Turbo code with $K=6144$ at a FER of $10^{-5}$ for both $R=\frac{1}{3}$ and $R=\frac{1}{2}$. We note, however, that at higher FERs the LTE Turbo code has significantly better performance than the polar codes. The polar codes only reach the performance of the LTE Turbo code at low FERs because the latter exhibits a relatively high error floor. Unfortunately, the polar code with $N=16384$ under BP decoding cannot reach the performance of the LTE Turbo code, even when a maximum of $I=30$ iterations are performed. We note that there exist Turbo codes that can even outperform the LTE Turbo code~\cite{Garzon2016}, thus increasing the potential gap in performance between polar codes and Turbo codes.

In Figure~\ref{fig:LTEHardwareequiFER}, we observe that, even though an SC decoder has the best hardware efficiency, \emph{on average} the SCL decoders have the best hardware efficiency among the polar decoders. Both the SC and BP decoders have very high area requirements when matching the FER performance of the LTE Turbo codes. We also observe that, on average, the LTE Turbo decoders have a similar hardware efficiency to the polar decoders.%

\section{Conclusion}
In this paper, we compared polar decoders both in terms of error-correction performance and hardware efficiency against LDPC and Turbo decoders for existing communications standards. Comparisons were made for the IEEE 802.11ad~\cite{IEEE802.11ad}, IEEE 802.11n~\cite{IEEE802.11n}, and IEEE 802.3an~\cite{IEEE802.3an}, and 3GPP LTE~\cite{LTE} communications standards. In most cases, BP and SC decoding are not powerful enough and more complex algorithms, such as SCL decoding, are needed in order to match the error-correction performance of LDPC or Turbo codes. Moreover, we have seen that the polar decoders that can match the error-correction performance of LDPC and Turbo codes usually have lower hardware efficiency than their LDPC and Turbo decoder counterparts. The low hardware efficiency stems mainly from the low throughput that these decoders achieve, and not so much from their area requirements. In conclusion, while significant improvements have been achieved over the past few years in the polar decoding literature, further work is required in order to match and surpass existing channel coding solutions. In particular, the direction of increasing the throughput of SCL decoders seems promising, since SCL decoders have the lowest area requirements and generally the best hardware efficiency out of the polar decoders in all comparisons of this paper.

\section*{Acknowledgement}
The authors would like to thank Huawei Technologies for financial support.

\balance
\begingroup
\raggedright
\bibliographystyle{IEEEtran}
\bibliography{IEEEabrv,bibliography}
\endgroup

\end{document}